\documentclass[%
 reprint,
superscriptaddress,
 amsmath,amssymb,
 aps,
]{revtex4-2}

\usepackage{graphicx}
\usepackage{dcolumn}
\usepackage{bm}
\usepackage{siunitx}
\usepackage{color}

\begin{document}

\preprint{APS/123-QED}

\title{Radically Uniform Spike Trains in  Optically Injected  Quantum Cascade Oscillators}

\author{Yibo Peng}
\affiliation{%
 School of Information Science and Technology, ShanghaiTech University, Shanghai 201210, China
}%
\affiliation{Shanghai Institute of Microsystem and Information Technology, Chinese Academy of Sciences, Shanghai 200050, China}
\affiliation{University of Chinese Academy of Sciences, Beijing 100049, China}
\author{Siting Liu}
\affiliation{%
 School of Information Science and Technology, ShanghaiTech University, Shanghai 201210, China
}%
\affiliation{%
 Shanghai Engineering Research Center of Energy Efficient and Custom AI IC, ShanghaiTech University, Shanghai 201210, China
}%
\author{Vassilios Kovanis}
\thanks{Corresponding author\\vkovanis@vt.edu}%
\affiliation{
Bradley Department of Electrical and Computer Engineering, Virginia Tech Research Center, Arlington, Virginia 22203, USA
}%

\author{Cheng Wang}
\thanks{Corresponding author\\wangcheng1@shanghaitech.edu.cn}%
\affiliation{%
 School of Information Science and Technology, ShanghaiTech University, Shanghai 201210, China
}%
\affiliation{%
 Shanghai Engineering Research Center of Energy Efficient and Custom AI IC, ShanghaiTech University, Shanghai 201210, China
}%

\begin{abstract}
It has been found that noise-induced excitability in quantum well and quantum dot  semiconductor laser systems usually produce spike patterns of non-uniform amplitude. In this letter,  we experimentally record  that an inter-subband quantum cascade laser injected with a monochromatic laser exhibits  a series of  highly-uniform spike trains in the time domain. Theoretical analysis demonstrates that such high uniformity has its origin in the ultrashort carrier lifetime of the quantum cascade laser gain medium that is  typically close to one picosecond.

\end{abstract}

\maketitle

Excitability is a common phenomenon in various nonlinear dynamical systems, including biological neurons, chemical reactions, and optical systems \cite{Izhikevich2007,IzhikevichTrans04}. When an excitable system is stimulated by a superthreshold perturbation, one or several spikes are created \cite{Wieczorek2002}. This process is followed by a refractory period, before which the system can not be excited again \cite{Garbin2017}. In the phase space, the excitable system undergoes a large excursion from the stable equilibrium \cite{Krauskopf2003}. Once the spike is fired, the system settles back to the equilibrium state.
On the other hand, a subthreshold perturbation can only trigger a small response without producing any giant spikes. Semiconductor lasers have been found to show excitability, when subjected to an external control including  saturable absorption \cite{Barbay2011}, optical feedback \cite{Giudici1997}, and the most used optical injection \cite{Goulding2007,Dillane2019}. Optical injection unidirectionally injects the monochromatic  beam of light of a master laser into a slave laser with a similar lasing frequency \cite{Valagiannopoulos2021,Himona2022,Herrera2022}. The slave diode laser is synchronized with the master laser within the stable regime, which is bounded by the Hopf bifurcation and the saddle-node bifurcation \cite{Ohtsubo2012}. Spiking pulses are usually observed in the vicinity of the saddle-node bifurcation, due to the existence of homoclinic bifurcation \cite{Wieczorek2002}. The stimulus of noise randomly perturbs the excitable laser systems and usually triggers a train of spikes \cite{Kelleher2011,Kelleher2011_2}. The amplitudes of these noise-induced spikes are usually irregular and vary a lot from one to another. On the other hand, finely controlled excitation of spikes is valuable for developing spike-based neuromorphic computing systems, which are attracting a lot of interest in recent years \cite{Turconi2013l,Prucnal2016,Brunner2018,Robertson2019}.

Investigations of excitability are mostly based on near-infrared interband semiconductor lasers, where the carrier lifetime is on the sub-nanosecond scale. In contrast, the laser emission of mid-infrared and terahertz quantum cascade lasers (QCLs) relies on the inter-subband transition. The carrier lifetime of QCLs is around one picosecond, which is two to three orders of magnitude smaller than common interband lasers \cite{Faist1994}. Due to the ultrashort carrier lifetime, QCLs have shown high stability with normal optical feedback, while complex nonlinear dynamics became rare events \cite{ZhaoBB2020,WangXG2020}. Nevertheless, our previous work found that optical feedback with a tilted angle could destabilize QCLs and trigger periodic oscillations, aperiodic oscillations, and low-frequency oscillations  \cite{WangXG2021,ZhaoBB2021}. In addition, current modulation of QCLs with optical feedback can stimulate low-frequency fluctuations and extreme pulses \cite{Jumpertz2016,Spitz2020}. When subject to optical injection, previous reports have theoretically identified both the Hopf bifurcation and the saddle-node bifurcation of QCLs \cite{Wang2013,Erneux2013}. Outside the stable locking regime, our recent work demonstrated that QCLs mostly produced periodic oscillations \cite{Peng2022}. In this work, we show that a QCL is excitable in the vicinity of the saddle-node bifurcation. It is strikingly found that the noise-induced spikes in the QCL exhibit highly-uniform amplitude, which is in contrast to those in interband semiconductor lasers. Theoretical modeling proves that the high uniformity of the spike amplitudes inherently originates from the ultrashort carrier lifetime of QCLs. 

\begin{figure}[htb]
\includegraphics[width=8.6cm]{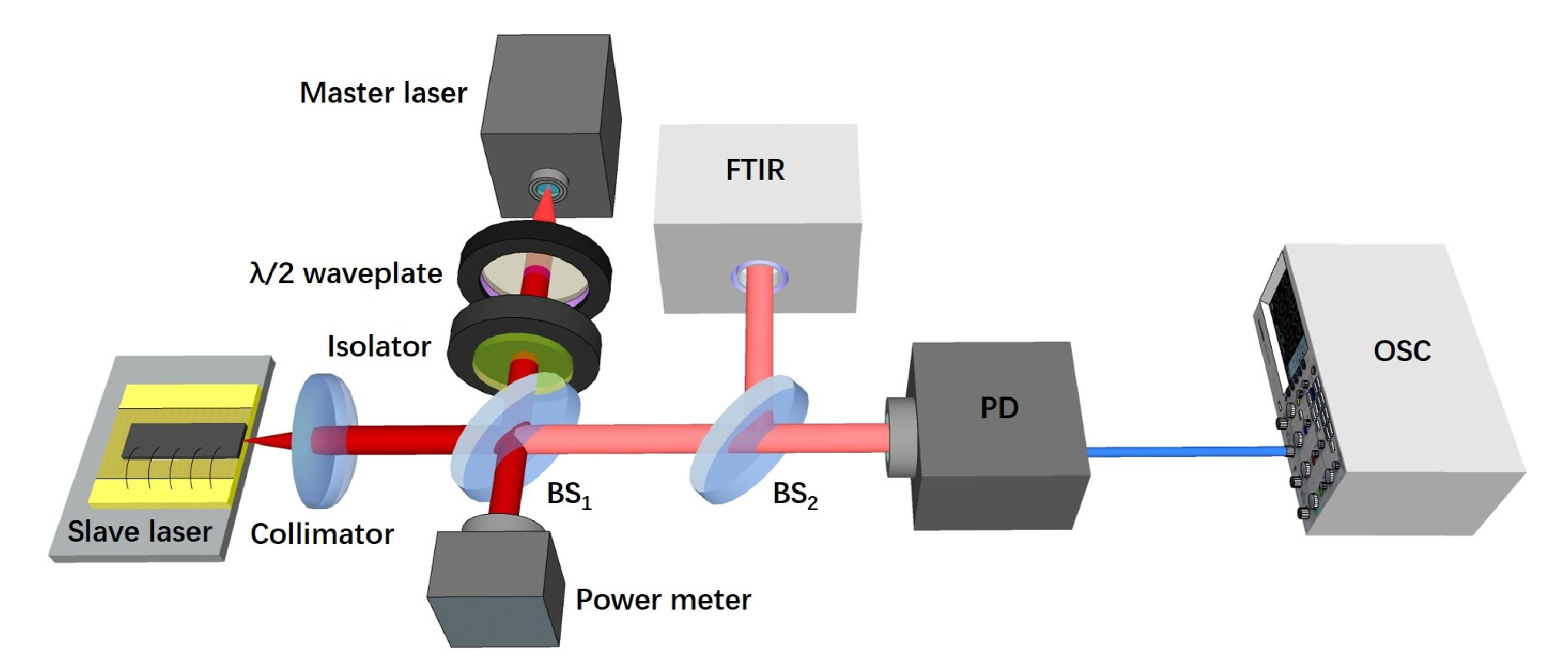}
\caption{Experimental tabletop setup of the QCL injected optically with monochromatic radiation. BS: beam splitter; FTIR: Fourier transform infrared spectrometer, that has  a resolution of 0.08 /cm; PD: photodetector; OSC: ultra oscilloscope.}
\end{figure}

Figure 1 depicts the experimental setup for the QCL subject to the monochromatic optical injection. The slave laser is a commercial distributed feedback QCL (Thorlabs), that is driven by a continuous-wave electronic current source. The operation temperature is maintained at 20 °C by using a thermoelectric cooler. The master laser is a tunable external-cavity QCL (Daylight solutions). The unidirectional injection is achieved through a polarization-dependent isolator. The injection strength is controlled by a half-wave plate and is monitored by a power meter. The optical spectrum is measured by a Fourier transform infrared spectrometer (FTIR, Bruker) with a resolution of 0.08 /cm.  The optical signal is converted to the electrical one through a HgCdTe photodetector (PD, Vigo) with a detection bandwidth of 560 MHz. The temporal waveform is recorded on a digital oscilloscope (OSC, 59 GHz bandwidth), and the sampling rate is fixed at 5.0 GSample/s. The free-running slave laser exhibits a lasing threshold of $I_{th}$=385 mA with an emission wavenumber around 2182 /cm. The optical frequency detuning  between the master laser and the slave laser is finely tuned by adjusting the pump current of the slave laser, instead of the master laser. The frequency tunability of the slave laser is measured to be about -770.5 MHz/mA.

\begin{figure}[htb]
\includegraphics[width=8.6cm]{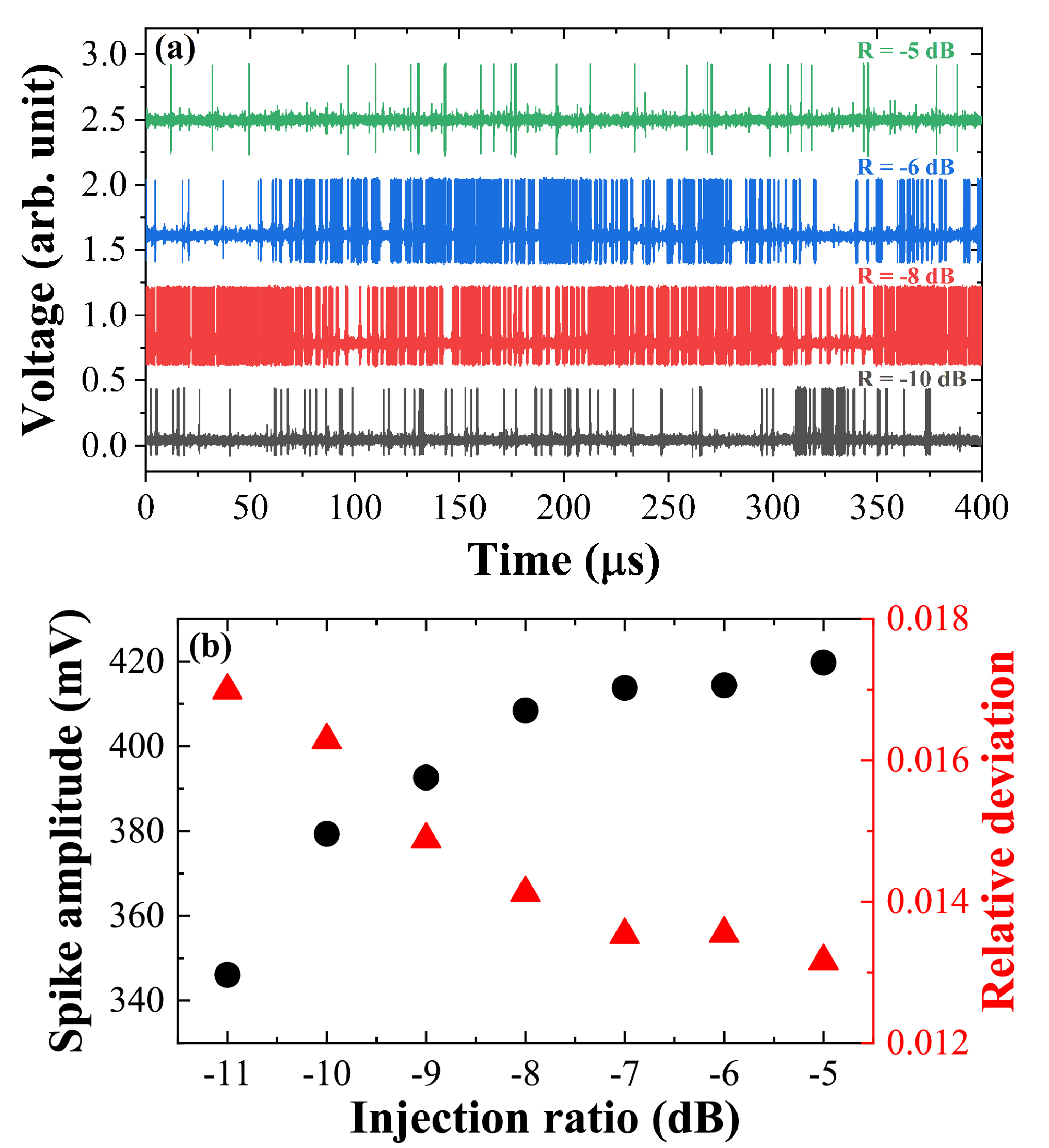}
\caption{(a) Spike trains of the QCL  recorded for several injection ratios. (b) Mean amplitude and relative standard deviation of the spikes as a function of the injection ratio.}
\end{figure}

When the slave QCL is pumped at 445 mA, the output power is 24.1 mW. The optical detuning frequency is -2.45 GHz, which is located in the vicinity of a saddle-node bifurcation. For injection ratios larger than -4 dB, the laser is locked in the stable, phase-locked  region and emits a continuous wave. When the injection ratio is reduced to the range of -4 to -11 dB, the QCL produces spikes as illustrated in Fig. 2(a). The spikes appear with random intervals, due to the random nature of noise excitation. Surprisingly, the spike amplitudes are highly uniform. This behavior is different from the noise-induced spikes of common interband semiconductor lasers, where the amplitudes vary substantially and distribute in a wide range \cite{Goulding2007}. Figure 2(a) also shows that the occurrence rate of spikes depends strongly  on the injection strength. An injection ratio of -8 dB triggers the most number of spikes compared to other injection powers. Figure 2(b) shows that the mean amplitude of the spikes recorded in a duration of 1 ms rises nonlinearly with increasing injection strength. Meanwhile, the relative standard deviation declines from 0.017 at the injection ratio of $R$=-11 dB down to 0.013 at $R$=-5 dB. Therefore, strong optical injection improves the uniformity of the spikes. 

\begin{figure}[htb]
\includegraphics[width=8.6cm]{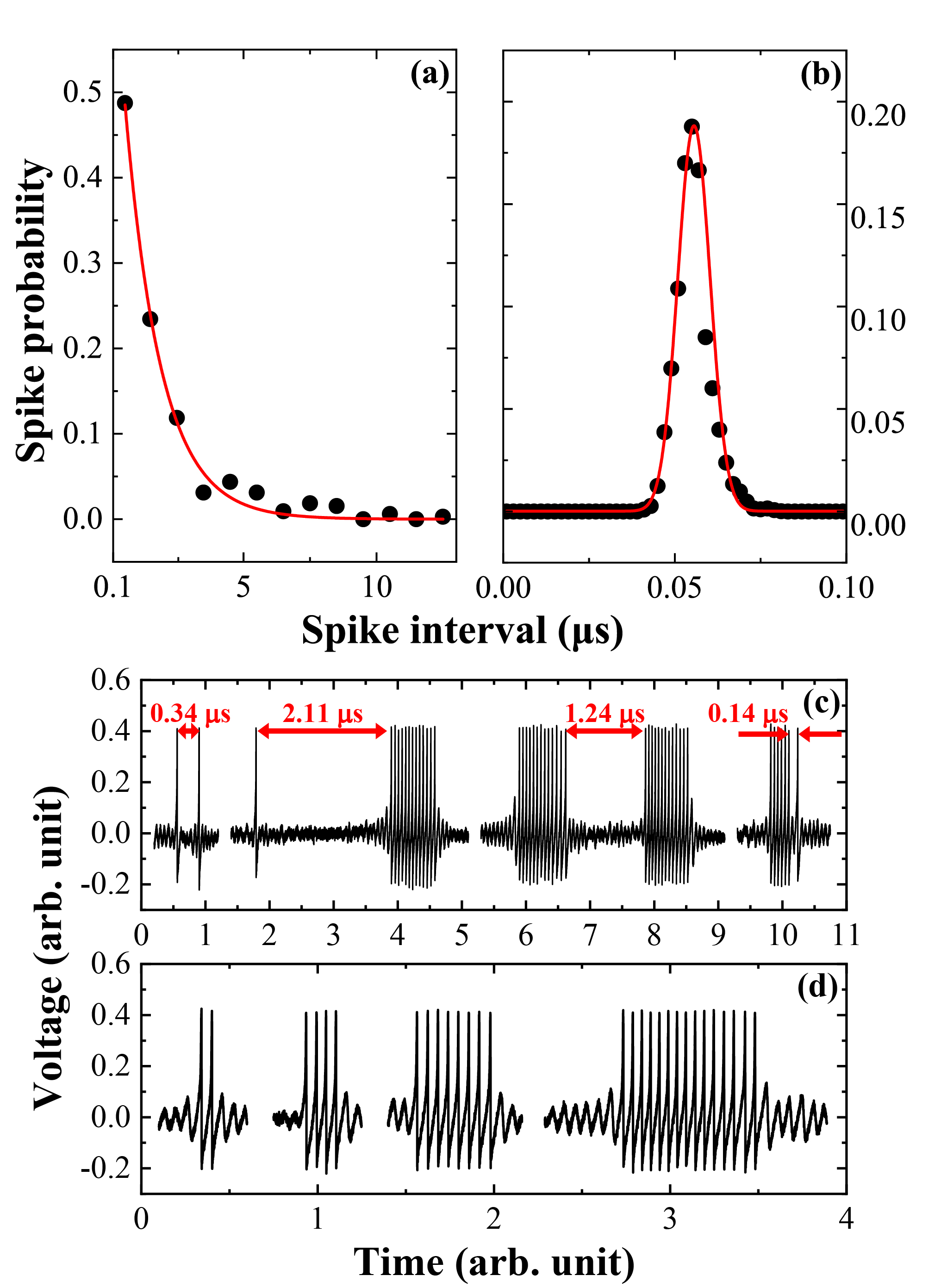}
\caption{Distribution of spikes for (a) intervals $>$ 0.1 $\si{\mu s}$ and (b) intervals $<$ 0.1 $\si{\mu s}$. (c) Waveforms counted in (a), and (d) waveforms counted in (b).}
\end{figure}

Figure 3 shows the distribution histogram of the spike intervals of the QCL, where the injection ratio is -6 dB and the analyzed time duration is 1 ms. For intervals above 0.1 $\si{\mu s}$, Fig. 3(a) shows that the distribution roughly follows the trend of exponential decay, which is a typical signature of noise-induced excitability \cite{Kelleher2011, Kramers1940}. The characteristic time of the decay is about 1.3 $\si{\mu s}$, which is governed by the noise strength. This characteristic time is more than one order of magnitude larger than that of common interband semiconductor lasers \cite{Kelleher2011}. Therefore, kicking QCLs out of the stable potential minimum is more difficult than the interband lasers. This behavior  happens because QCLs are intrinsically more stable than the latter, owing  to the ultrashort carrier lifetime and the small linewidth broadening factor. Indeed, the high stability of QCLs has been widely demonstrated via external perturbations from optical feedback \cite{ZhaoBB2020,WangXG2020}. Temporal waveform analysis in Fig. 3(c) shows that these intervals include three categories: spike-to-spike interval (0.34 $\si{\mu s}$), spike-to-burst interval (2.11 $\si{\mu s}$ and 0.14 $\si{\mu s}$), and burst-to-burst interval (1.24 $\si{\mu s}$),  which are illustrated in Fig. 3(c). For intervals below 0.1 $\si{\mu s}$, Fig. 3(b) proves that the distribution roughly follows a Gaussian function, with a mean value of 56 ns and a standard deviation of 4.7 ns. Temporal waveform analysis in Fig. 3(d) shows that this distribution comes from the intra-burst pulses, which are evenly spaced. Each burst consists of a different number of pulses, ranging from two up to more than ten.    

In order to dissect the physical mechanism for producing spikes of uniform amplitudes, we model the  optically injected QCLs using  a set of single-mode rate equations, and the effects of quantum noise on the photon and carrier dynamics are included as well \cite{WangXG2018}:

\begin{small}
\begin{align}
&\frac{d N_3}{d t} \! = \!\eta \frac{I}{q} \!- \!\frac{N_3}{\tau_{32}} \!- \!\frac{N_3}{\tau_{31}} \!- \!G_0 S \Delta N \!+ \!F_3(t) \\
&\frac{d N_2}{d t} \! = \!\frac{N_3}{\tau_{32}} \!- \!\frac{N_2}{\tau_{21}} \!+ \!G_0 S \Delta N \!+ \!F_2(t) \\
&\frac{d N_1}{d t} \! = \!\frac{N_3}{\tau_{31}} \!+ \!\frac{N_2}{\tau_{21}} \!- \!\frac{N_1}{\tau_{\text {out }}} \!+ \!F_1(t) \\
&\frac{d S}{d t} \! = \!\left( \!m G_0 \Delta N \!- \!\frac{1}{\tau_p} \!\right) \! S \!+ \!m \beta \frac{N_3}{\tau_{s p}} \!+ \!2 k_c \sqrt{S_{inj} S} \cos \phi+F_s(t) \\
&\frac{d \phi}{d t} \! = \!\frac{\alpha_H}{2} \! \left( \!m G_0 \Delta N \!- \!\frac{1}{\tau_p} \!\right) \!- \!2 \pi \Delta F_{inj} \!- \!k_c \sqrt{\frac{S_{inj}}{S}} \sin \phi \!+ \!F_\phi(t)
\end{align}
\end{small}

\noindent
where $N_{3,2,1}$ are carrier numbers in the upper lasing level, the lower lasing level, and the bottom level, respectively. $S$ is the photon number in the injection-locked laser cavity, and $\phi$ is the phase difference between the slave laser and the master laser. $I$ is the pump current, $\eta$ is the current injection efficiency, $G_0$ is the gain coefficient ($G_0 \!= \!5.3 \! \times \! 10^4$ /s), $m$ is the number of gain stage ($m$=30), $\beta$ is the spontaneous emission factor ($\beta \!= \!10^{-6}$), and $\Delta N$ is the population inversion expressed as $\Delta N \! = \! N_3 \! - \! N_2$. $S_{inj}$ is the photon number of the master laser, and the injection ratio is defined as $R_{inj} \! = \! S_{inj} /S_0$, with $S_0$ being the photon number of the free-running slave laser. $\Delta F_{inj}$ is the optical detuning frequency between the master laser and the slave laser, and $k_c$ is the coupling coefficient between the two lasers. The carrier lifetimes for each level are $\tau _{32}$=2.0 ps, $\tau _{31}$=2.4 ps, $\tau _{21}$=0.5 ps, $\tau _{out}$=0.5 ps. The photon lifetime is $\tau _p$=3.7 ps, and the spontaneous emission lifetime is 7.0 ns. The carrier noise $F_{3,2,1}$ and the spontaneous emission noise, quantum noise $F_{S,\phi}$ are modeled by the Langevin noise sources, respectively \cite{Coldren2012}. Detailed correlation relations between the noise sources refer to \cite{WangXG2018}. The linewidth broadening factor is fixed at $\alpha _H$=-0.5. We note that measured linewidth broadening factors of QCLs in the literature range from -1.0 up to more than 2.0 \cite{ZhaoBB2020,OpacakOptica21}. However, the optical injection can reduce its value due to the reduction of carrier population \cite{WangOE15}.

\begin{figure}[htb]
\includegraphics[width=8.6cm]{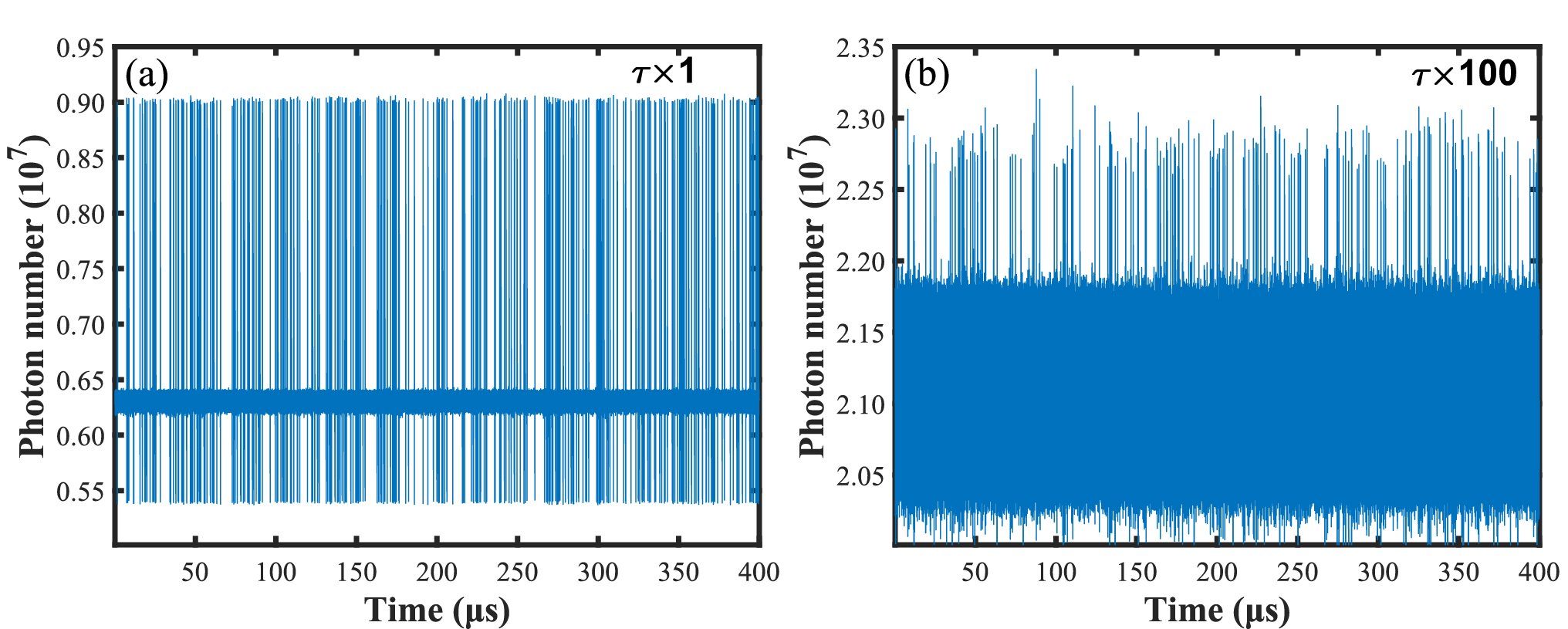}
\caption{Simulated spike trains of (a) the common QCL with ultrashort carrier lifetimes and (b) the artificial QCL with 100-time longer carrier lifetimes. The  detuning frequency is -2.1 GHz in (a) and -1.2 GHz in (b), respectively. The injection ratio is -5 dB for both cases.}
\end{figure}

In the vicinity of the saddle-node bifurcation, the simulation in Fig. 4(a) successfully reproduces the spike trains with highly uniform amplitudes, which agrees well with the experimental results in Fig. 2(a). The relative standard deviation of the simulated spikes is 0.007. This deviation is smaller than the experimental one in Fig. 2(b). It is likely due to the fact that the rate equation model does not take into account technical noise sources, including the current source noise, the operation temperature fluctuation, and the mechanical vibration. Once we remove the noise sources in the rate equation model, the spiking dynamics do not occur. Therefore, it is verified that the generation of spikes in QCLs arises from the noise-induced excitability.  In order to explore the mechanism governing the uniformity of amplitudes, we artificially enlarge all the carrier lifetimes by a factor of 100, while other parameters remain unchanged. That is, the carrier lifetime of the QCL becomes about 100 ps. Figure 4(b) shows the spike trains of this artificial QCL with optical injection. Interestingly, the spike amplitudes are not uniform anymore but vary significantly from spike to spike. The corresponding relative standard deviation reaches as high as 0.07, which is one order of magnitude larger than the one with ultrashort carrier lifetimes in Fig. 4(a). These non-uniform spikes are similar to those commonly observed in interband semiconductor lasers with a lifetime of sub-nanosecond \cite{Goulding2007}. Therefore, we can conclude that the high uniformity of the spikes in QCLs is owing to the ultrashort carrier lifetime. Indeed, the fast gain recovery process in the QCL gain medium is helpful to store strong enough energy for each firing spike, whenever the noise stimulus perturbs the laser system \cite{Choi2008,Green2009}.

We remark that the uniform spikes occur in a broad regime (bounded by the injection ratio and the detuning frequency) in the vicinity of the saddle-node bifurcation.  In addition, we have observed uniform spikes for the pump currents both close to and well above the lasing threshold of the QCL. Therefore, the excitation of uniform spikes in QCLs with optical injection is robust and reliable. These uniform spikes are highly valuable for developing neuromorphic computing systems or spiking neural networks. These networks consist of many spiking neurons and the neuron dynamics are usually described by the popular leaky-integrate-and-fire (LIF) model \cite{Izhikevich2007}. In this model, the membrane potential of the neuron continuously experiences an exponential decay toward the resting value. Once the neuron receives a spike, the membrane potential increases by an amount proportional to the amplitude of the spike. When the potential reaches a certain threshold, the neuron fires a spike and resets its membrane potential. With this repeated process, the neuron generates a series of spikes or a spike train. To perform neuromorphic computing using these neurons, the spike train is usually encoded by the spiking rate (rate coding) or the spiking time (temporal coding) \cite{Gerstner02}. To successfully code the message, it is crucial to ensure the uniformity of the spikes, i.e., all the input spikes and output spikes must have identical amplitude. Otherwise, a large input spike may incorrectly lead to the fast rate or early appearance of the output spike, and vice versa.
Consequently, an error occurs due to the change of the message encoded on the output spike train. Indeed, CMOS-based electronic neuromorphic computing systems usually own uniform spike trains \cite{SroujiAPLP22,ChalkiadakisPRE22}. However, the spikes produced from near-infrared lasers like vertical-cavity surface-emitting lasers are usually not uniform, which prevents high-quality implementation of the LIF model based on laser diodes \cite{Robertson2019,Owen2022}.

In summary, we have experimentally recorded the observation of highly uniform spike trains in a tabletop optically injected quantum cascade system. The spike trains are excited via quantum noise, in the vicinity of a saddle-node bifurcation, where a collision and disappearance of two equilibria points are happening. Our theoretical analysis demonstrates that such high uniformity is born from the ultra-short carrier lifetime of the optical gain medium, which is more than two orders of magnitude smaller than that of common quantum well  laser diodes. We assert  that uniform spikes are of prime importance for developing next-generation photonic neuromorphic computing systems. Future work will investigate the detailed characteristics of spikes excited by properly controlled input signals instead of just quantum noise, such as the excitation threshold, the spike latency, and the refractory period.  

This work was financially supported by Shanghai Natural Science Foundation (20ZR1436500) and the work of VK was supported via generous  gifts  to the Virginia Tech Foundation.


\bibliography{reference}
\end{document}